\documentclass[twocolumn,prl,superscriptaddress,floatfix,showpacs]{revtex4-1}
\usepackage[utf8]{inputenc}
\usepackage{amsmath}
\usepackage{amssymb}
\usepackage{graphicx}

\usepackage{graphics}
\usepackage{epsfig}

\newcommand{\be}{\begin{equation}} \newcommand{\ee}{\end{equation}}
\newcommand{\bea}{\begin{eqnarray}} \newcommand{\eea}{\end{eqnarray}}

\newtheorem{proposition}{Proposition}
\newtheorem{definition}{Definition}

\begin{document}

\title{Percolation in Media with Columnar Disorder}

\author{Peter Grassberger}
\affiliation{JSC, FZ J\"ulich, D-52425 J\"ulich, Germany}

\author{Marcelo R.\ Hil\'ario}
\affiliation{Dep.\ of Mathematics, UFMG, 30161-970 Belo Horizonte, Brazil}

\author{Vladas Sidoravicius}
\affiliation{Courant Institute of Mathematical Sciences, NYU, New York, USA.}
\affiliation{NYU-ECNU Institute of Mathematical Sciences at NYU Shanghai, China}
\affiliation{Cemaden, S\~ao Jos\'e dos Campos, Brazil}

\date{\today}
\begin{abstract}
We study a generalization of site percolation on a simple cubic lattice, where 
not only single sites are removed randomly, but also entire parallel columns of sites. We show 
that typical clusters near the percolation 
transition are very anisotropic, with different scaling exponents for the sizes parallel and perpendicular
to the columns. Below the critical point there is a Griffiths phase where cluster size distributions and spanning
probabilities in the direction parallel to the columns have power law tails with continuously varying 
non-universal powers. This region is very similar to the Griffiths phase in subcritical directed percolation
with frozen disorder in the preferred direction, and the proof follows essentially the same 
arguments as in that case. But in contrast to directed percolation in disordered media, the number of active 
(``growth") sites in a growing cluster at criticality shows a power law, while the probability of a cluster 
to continue to grow shows logarithmic behavior.
\end{abstract}
\maketitle

\section{Introduction}

The percolation transition is continuous and in a unique and well understood universality class called 
``ordinary percolation" (OP). This is at least the folklore and what textbooks say 
\cite{Stauffer}. Although it is also true in many cases, it is not always true and reality is more complex --
even if we disregard such well known problems like directed \cite{Hinrichsen}, rigidity \cite{Moukarzel}, 
and bootstrap \cite{Adler} percolation.

In some of the unusual percolation scenarios the rules by which clusters grow are modified, as e.g.\ in 
explosive percolation \cite{Achlioptas}, cooperative infection \cite{Dodds,Janssen,Bizhani} and the closely
related $k$-core percolation \cite{Goltsev,Baxter}, `agglomerative percolation' \cite{Christensen,Lau},
percolation in multiplex networks if connectivity is demanded for each layer \cite{Buldyrev,Son}, or 
co-infections \cite{Cai}. The fact that this can lead to different behavior might not be so surprising.

Much more surprising are situations where the basic rules of cluster growth and connectivity establishment
are unchanged, and the anomalous behavior results only from the particular geometry of the underlying lattice 
or network. These include percolation in media with long range correlations \cite{Schrenk0}, pacman and interlacements percolation \cite{Abete,Sznitman0,Sznitman} and `drilling 
percolation' \cite{Kantor,Hilario,Schrenk,Grassberger2016,HS}, where the point defects appearing in Bernoulli site percolation are replaced by removed (`drilled') entire columns of sites. But more spectacular is the model 
of growing random networks of Callaway {\it et al.} \cite{Callaway} that shows a Kosterlitz-Thouless (KT) 
transition, and the very old model of 1-d percolation with long range links \cite{Aizenman} that shows a 
KT-like transition that is indeed discontinuous \cite{Grassberger2013}. The latter model is closely related to 
percolation on some hierarchical structures \cite{Boettcher}.

In the present paper we study a variant of 3-d site percolation on the simple cubic lattice with less exotic 
properties, but yet with some surprising changes compared to OP. In this model sites are 
removed by two superimposed but independent mechanisms:

(i) As in OP, we remove single sites ${\bf x} = (x,y,z)$ with probability $1-p_B$ (here 
``B" stands for `Bernoulli').

(ii) In addition, we remove entire columns of sites $C(x,y) = \{(x,y,z),  0 < z \leq L_z\}$ parallel to the 
z-axis with probability $1-p_z$. Here, it is assumed that the lattice is of size $L\times L\times L_z$.

Due to the removal of columns, the clusters are of course less isotropic than in ordinary
percolation. As we shall see, they are elongated in the z-direction. In this respect the model resembles 
directed percolation, but notice that there is no bias towards the positive or negative z-direction.

This model is also similar to the ``drilling percolation" model where no single sites are removed and columns are drilled not only parallel to the z-axis, but with the same probability parallel to the x- and y-axes (see \cite{Schrenk, HS} for recent studies). In  drilling percolation, typical critical clusters are still cigar shaped but they are oriented along any one of the three axes. 
Take a cluster that happens to be oriented along the z-axis. For that cluster
it is obviously crucial that part of the randomness is columnar in the z-direction, while the fact that 
the other defects are also columnar is presumably much less important. From the point of such
a cluster, our present model can be seen as a simplification of drilling percolation where the ``perpendicular"
columns are replaced by points. As we shall see, this makes not only the numerics much more distinct, but it
simplifies also the mathematical treatment.
Also mixed Bernoulli and drilling percolation appear naturally when studying drilling percolation in $4$ dimension or higher.
Indeed, the restriction of $4$-d drilling percolation to a $3$-d sublattice gives rise to a $3$-d drilling percolation superposed by a Bernoulli site percolation (see \cite{HS}).

\section{Phase diagram}

In this and the following two sections we shall present numerical results for clusters that grow a finite 
time from point seeds according to a modified Leath algorithm. Lattices are so big that clusters never touch the boundary (which is explicitly 
checked), thus there are no finite (lattice-)size corrections. 

Consider first the case $p_B=1$, i.e. only columns are removed and no single points. In this limit, the 
problem reduces to 2-d site percolation. Thus clusters can be extended in the $x$ and $y$ direction 
only for $p_z > p^{(2)}_c = 0.592746\ldots$. Clusters are infinite also for $p_z < p^{(2)}_c$,
but they are then strictly 1-dimensional \footnote{Strictly spoken this is not true, since the cluster mass distribution for 2-d percolation has tails even below the critical point, but we shall assume that these tails do not affect the phase diagram.}.

\begin{figure}
\begin{centering}
\includegraphics[scale=0.30]{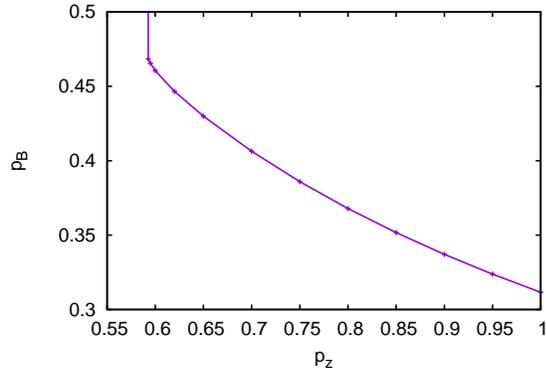}
\par\end{centering}
\caption{\label{fig1} (color online)  Critical line for the model with columns removed with probability $1-p_z$, and single points removed with probability $1-p_B$. Percolating clusters exist above the curve. The left end of the curve is at $p_z = p^{(2)}_c$, and the right end at $p_z=1$ has $p_B = p^{(3)}_c$.}
\end{figure}

When we now turn to $p_B < 1$, clusters at $p_z < p^{(2)}_c$ will no longer extend infinitely far in the 
z-direction, but their length distribution will be cut off.
Also it is clear that, for any $p_B < p_c^{(3)}$ all cluster are finite regardless of the value of $p_z$. 
Therefore, for a fixed value $p_B$ in $]p_c^{(3)}, 1[$, there exists a critical value for $p_z$
\be
    p_{z,c} = p_{z,c}(p_B) > p^{(2)}_c,
\ee
or equivalently, for a fixed value any fixed $p_z > p_c^{(2)}$, there exists
\be
    p_{B,c} = p_{B,c}(p_z).
\ee
When, finally, $p_z=1$ (i.e., no columns are drilled at all), then $p_{B,c}(1) = p^{(3)}_c = 0.311607\ldots$.

\begin{figure}
\begin{centering}
\includegraphics[scale=0.30]{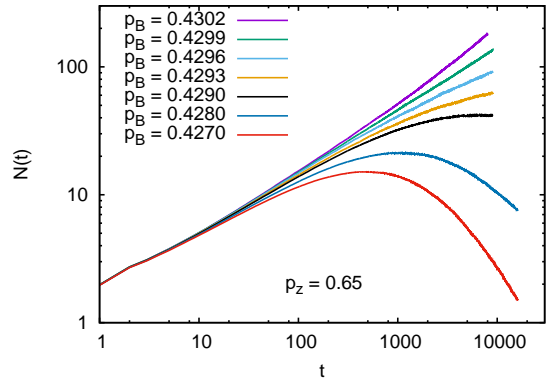}
\par\end{centering}
\caption{\label{fig2} (color online)  Average number of growth sites versus time $t$, for $p_z = 0.65$. The 
critical point for this $p_z$ is approximately $p_{B,c} = 0.4299$.}
\end{figure}

The phase boundary obtained by numerical simulations is shown in Fig.~1. In obtaining it we let clusters
grow with a modified Leath algorithm from point seeds, for typically $5\times 10^3$ to $10^4$ time steps, 
and observe the average number $N(t)$ of ``growth sites", i.e. of newly wetted sites. At the percolation
threshold we expect a power law
\be
    N(t) \sim t^\mu.       \label{mu}
\ee
Alternatively, if we take directed percolation in media with frozen disorder \cite{Moreira,Cafiero} as a guide, we 
might expect logarithmic scaling. As suggested by Fig.~2, the power law scaling at criticality seems to be 
correct, at least for $p_z \geq 0.65$. The numerical value of the exponent $\mu$ and its dependence on $p_z$ 
will be discussed later.

Notice that the critical curve does not continuously rise up to $p_{B,c} =1$ as $p_z$ approaches the 2-d 
critical point, but jumps to 1 from a finite value which is strictly smaller than $p_c^{(2)}$. Although this is somewhat unexpected, it can easily be 
understood. For any $p_z > p^{(2)}_c$ the removal of 
columns leaves a connected region whose intersection with any plane $z={\rm const}$ contains with non-zero
probability an infinite cluster and thus also an infinite path. This path corresponds in 3-d space to a 
crumpled 2-d sheet. On this sheet, percolation occurs for any $p_B > p^{(2)}_c$. Thus the curve in Fig.~1 must 
end at $p_z = p^{(2)}_c$ at a value $p_B \leq p^{(2)}_c$. It might be that the slope 
of the critical curve diverges in this limit, but more precise simulations would be necessary to settle this
question.

The above argument does not explain why the critical curve approaches a value strictly smaller than $p_c^{(2)}$ as $p_z \searrow p_c^{(2)}$.
A possible explanation for this, that we do not make completely rigorous here is that, at $p_z=p_c$ we would consider the crumpled $2$-d sheet supported by the backbone of the incipient infinite cluster.
The existence of finite clusters in $2$-d emerging from the backbone guarantees that in $3$-d there are columnar structures adjacent to the sheet that enhance the connectivity, lowering thereby the critical point.

\section{Cluster shapes}

During the same runs we also measured the longitudinal and transversal r.m.s sizes 
$R_z(t) = [\langle z^2\rangle]_t^{1/2}$ and $R_{xy}(t) = [\langle (x^2+y^2)\rangle/2]_t^{1/2}$,
where the averages go over all growth sites at time $t$. For $p_z=1$ (i.e. for ordinary percolation) 
they are of course the same, but for $p_z<1$ they are clearly different. Typical results, again for 
$p_z = 0.65$, are shown in Fig.~3.

\begin{figure}
\begin{centering}
\includegraphics[scale=0.30]{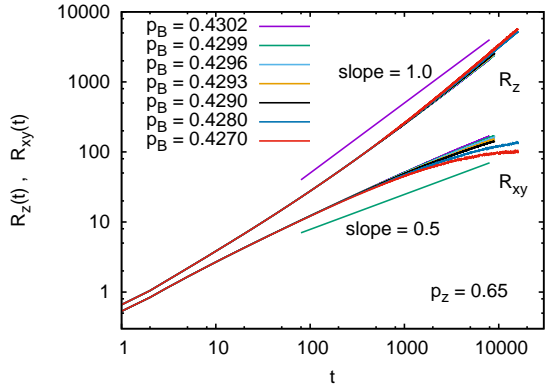}
\par\end{centering}
\caption{\label{fig3} (color online)  Root-mean-square longitudinal and perpendicular sizes as functions of $t$, 
for the same values of $p_z$ and $p_B$ used also in Fig.~2. The straight lines are drawn for comparison with
the power laws $R_z(t) \sim t$ and $R_{xy}(t) \sim \sqrt{t}$.}
\end{figure}

We see a dramatic asymmetry. For large critical clusters it seems that 
\be
   R_z(t) \sim t, \quad R_{xy}(t) \sim t^{1/2},    \label{R_t}
\ee
but there are huge corrections. Indeed, $R_z(t)$ seems to increase for large $t$ faster than $\sim t$, which 
can of course not be the asymptotic behavior. Rather, the data can be explained qualitatively by the following
scenario: For small $t$ the clusters grow roughly spherical, with $R_z(t)$ only slightly larger than $R_{xy}(t)$.
But the lateral growth nearly stops after some time (lateral growth alone would be subcritical), and for larger $t$
the growth is mostly longitudinal, in regions not containing any removed columns. At the transition between
these two regimes the growth sites on the spherical periphery of the cluster die and are replaced by growth
sites at the two ``end caps", leading thus to a faster than linear growth of $R_z(t)$ in the transition region.

Indeed, this effect is even more pronounced for $p_B < p_{B,c}$, since there the lateral growth is even sharper
cut off.

\section{Dependence on $p_z$}

Since the asymmetry decreases with increasing $p_z$, there is of course no chance to verify Eq.~(\ref{R_t}) 
numerically for $p_z$ close to 1. On the other hand, it is clear from the data that even for $p_z = 0.95$ the ratio 
$R_z(t)/R_{xy}(t)$ does not tend asymptotically towards a constant, suggesting that they satisfy different scaling
laws. The simplest assumption is that Eq.~(\ref{R_t}) holds for all $p_z < 1$. This is indeed suggested by 
all data for $p_z < 0.7$.

\begin{figure}
\begin{centering}
\includegraphics[scale=0.30]{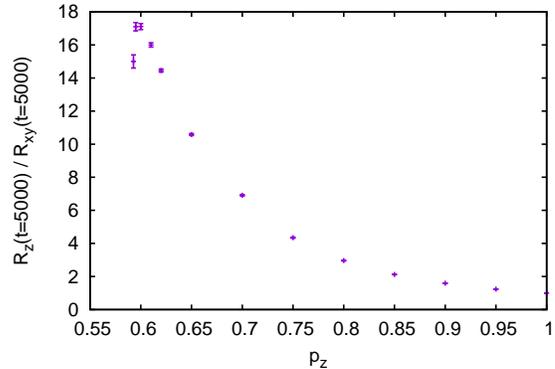}
\par\end{centering}
\caption{\label{fig4} (color online)  Ratios between longitudinal and perpendicular sizes for $t = 5000$ at 
the critical curve, plotted against $p_z$.}
\end{figure}

Nevertheless there are indications that the critical behavior might change somewhere between $p_z = p^{(2)}_c$ 
and $p_z = 0.62$. The first hint is that $R_z(t)/R_{xy}(t)$ for fixed $t$ does not increase monotonically with 
decreasing $p_z$, as seen from Fig.~4 where $R_z(t)/R_{xy}(t)$ for $t=5000$ is plotted against $p_z$. There 
seems to be a maximum at $p_z \approx 0.60$. 

\begin{figure}
\begin{centering}
\includegraphics[scale=0.30]{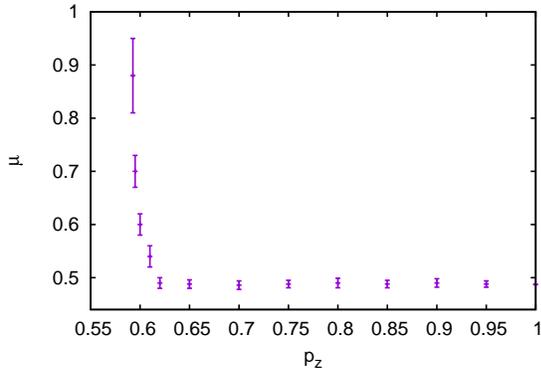}
\par\end{centering}
\caption{\label{fig5} (color online)  Growth exponent $\mu$ (defined in Eq.~(\ref{mu})) plotted against $p_z$.}
\end{figure}

A stronger (but still not convincing) indication is given by the dependence of the exponent $\mu$ on $p_z$, 
shown in Fig.~5. For all $p_z \geq 0.62$ it is within errors equal to 0.487, the value for OP \cite{Wang}.
But for smaller $p_z$ it seems to increase steeply, reaching finally a value that is clearly different.

\begin{figure}
\begin{centering}
\includegraphics[scale=0.30]{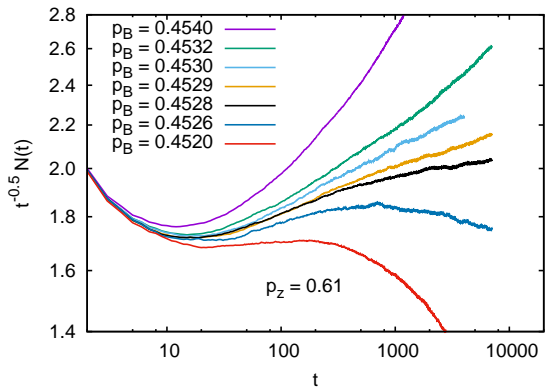}
\par\end{centering}
\caption{\label{fig6} (color online)  Average number of growth sites versus time $t$, for $p_z = 0.61$. To 
make the plot more significant, the actual variable plotted is $t^{-0.5} N(t)$.}
\end{figure}

We have not seen other qualitative changes of critical clusters near $p_z \approx 0.60$ to 0.62, whence  
the occurrence of a (tri-) critical point on the phase boundary shown in Fig.~1 would be rather puzzling.
In view of this we propose a different scenario, where actually $\mu = 0.487$ holds for all $p_z > p^{(2)}_c$, 
$R_z(t)/R_{xy}(t)$ for fixed $t$ increases for all $p_z > p^{(2)}_c$, and where only the behavior exactly 
at $p_z = p^{(2)}_c$ is different. The deviations seen for $p_z < 0.62$ would then be just cross-overs.
This alternative scenario is supported by Fig.~6, which shows $t^{-0.5} N(t)$ for $p_z = 0.61$. At first,
it seems that the critical point is $p_{B,c} = 0.4530$, since that curve seems to be most straight for large $t$.
This would give $\mu = 0.54$ as used in Fig.~5. But a closer look indicates that all curves for $0.4529 < p_B 
< 0.4532$ are slightly bent upwards for $t > 1000$, indicating that the critical value is indeed $p_{B,c} \leq 0.4528$,
implying that $\mu $ is much smaller and compatible with the universal value 0.487. The same behavior is also
seen for $p_z = 0.60$, but much better data would be needed for an unambiguous decision between the two scenarios.
\begin{figure}

\begin{centering}
\includegraphics[scale=0.30]{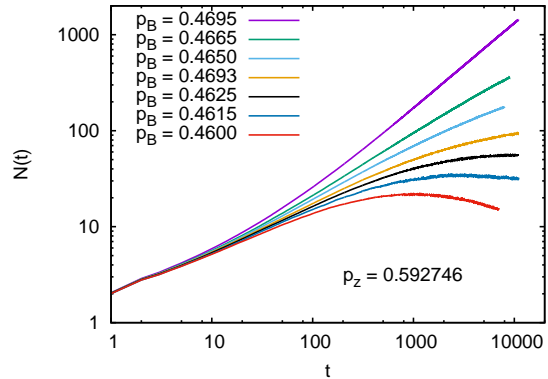}
\vglue -1.cm
\includegraphics[scale=0.30]{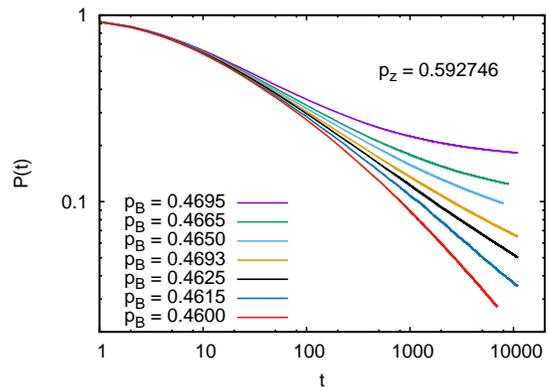}
\par\end{centering}
\caption{\label{fig7} (color online) Log-log plots of $N(t)$ (upper panel) and $P(t)$ (lower panel) for 
   $p_z = p^{(2)}_c = 0.592746$. Notice that there is no value of $p_B$ where both $N(t)$ and $P(t)$ are 
   described either by power laws or by logarithmic $t$-dependence.}
\end{figure}

Finally we should point out that the survival probability $P(t)$ does not satisfy a power law at criticality.
To demonstrate this we show in Fig.~7 both $N(t)$ (panel a) and $P(t)$ (panel b) for $p_z = 0.592746 = p^{(2)}_c$,
and for the same values of $p_B$. There is no value of $p_B$ for which both $N(t)$ and $P(t)$ show power laws.
Nor is there a value of $p_B$ for which both show logarithmic $t$-dependence.
From $N(t)$ our best estimate is $p_{B,c} = 0.4695(10)$, for which $P(t)$ decreases clearly much slower than a 
power of $t$.

\section{Griffiths phase}

\subsection{Spanning probabilities}

Let us now consider finite lattices of size $L\times L\times L_z$ with open boundary conditions laterally and 
with $L_z > L$ in general.
The base surface $z=0$ is assumed to be all wetted, and the growth is allowed to proceed only into the positive
cylinder $0 < x,y \leq L,\;\; z>0$. Spanning clusters exist on this lattice iff the growth continues until it reaches 
the upper surface $z = L_z$. Indeed for any height $h \leq L_z$ the spanning probability $P_{\rm span}(L,h;p_B,p_z)$ 
is exactly equal to the probability that the growth reaches height $h$.

A lower bound on this probability in the region $p_B \in ]p_c^{(3)}, p_{B,c}(p_z)]$ can be obtained for any 
fixed $p_z > p_c^{(2)}$ and sufficiently large $L$ as follows. Consider in the base surface a connected region 
$\cal{A}$ of area $A \leq L^2$, e.g.\ a square of size $\ell\times \ell$ with $\ell \leq L$. But the shape of 
$\cal{A}$ can be arbitrary, provided it is characterized by length scales much larger than the correlation length 
$\xi(p_B)$ of 3-d site percolation with $p=p_B$. In particular we can also take a rectangle $L\times \xi(p_B)$ 
or a strip of width $\xi(p_B)$ along one of the two diagonals, as considered in \cite{Schrenk}. We assume of 
course that $L \gg \xi(p_B)$.

Consider now instances where no column is drilled in $\cal{A}$. Since drilling is random with probability 
$1-p_z$ per site, the probability for this to happen is 
\be
    P_{\cal A} = e^{-(1-p_z)A}.   \label{PA}
\ee
The spanning probability is then
\be
    P_{\rm span}(L,L_z;p_B,p_z) \geq P_{\cal A} \times Q({\cal A},L_z,p_B)   \label{P_span}
\ee
where $Q({\cal A},h,p_B)$ is the probability that a cluster grown on the cylinder with base surface $\cal{A}$
grows up to height $L_z$.

The latter can be easily estimated for $L_z \gg L$, using the usual scaling picture for 3-d ordinary percolation. 
Notice that we have $p_B > p^{(3)}_c$, thus the correlation length $\xi = \xi(p_B)$ is finite and was assumed 
to be $\ll L$. In this case a cluster occupying initially the entire base surface will continue to grow until it
dies {\it simultaneously and independently} in all $A/\xi^2$ patches of diameter $\xi$. Thus the probability for 
such a cluster to die exactly at any height $h \gg 1$ scales as 
\be
   q_{\rm die} \sim  e^{-b A/\xi^2}  \label{q_die}
\ee 
(with $b$ being a constant of order 1), and 
\be
   Q({\cal A},L_z,p_B) \sim (1-q_{\rm die})^{L_z/\xi} = (1-e^{-b A/\xi^2})^{L_z/\xi},   \label{Q}
\ee
Combining Eqs.~(\ref{PA}) to (\ref{Q}), we obtain
\be
   -\ln P_{\rm span} \lesssim (1-p_z)A - (L_z/\xi) \ln (1-e^{-b A/\xi^2}).   \label{PP}
\ee
While the first term on the right hand side increases with $A$, the second one decreases. The minimum of the 
r.h.s. (and thus the upper bound on $P_{\rm span}$) is obtained by setting the derivative
with respect to $A$ equal to zero, which gives
\be
    e^{bA/\xi^2} \sim 1+\frac{L_z b}{(1-p_z)\xi^3},
\ee
which in turn gives for large $L_z$
\be
    A \approx c \ln L_z,   \label{A}
\ee
where the constant $c$ depends on $p_z$ and $p_B$.
Inserting this into Eq.~(\ref{PP}) gives finally a power law for large $L_z$
\be
    P_{\rm span}(L,L_z;p_B,p_z) > {\rm const} L_z^{-\alpha},  \label{Ps}
\ee
where the non-universal exponent $\alpha$ depends on $p_z$ and $p_B$. Equation (\ref{A}) implies that
this bound is true whenever $L > \ln L_z$, in particular for any fixed aspect ratio $L_z/L$.

\begin{figure}
\begin{centering}
\includegraphics[scale=0.30]{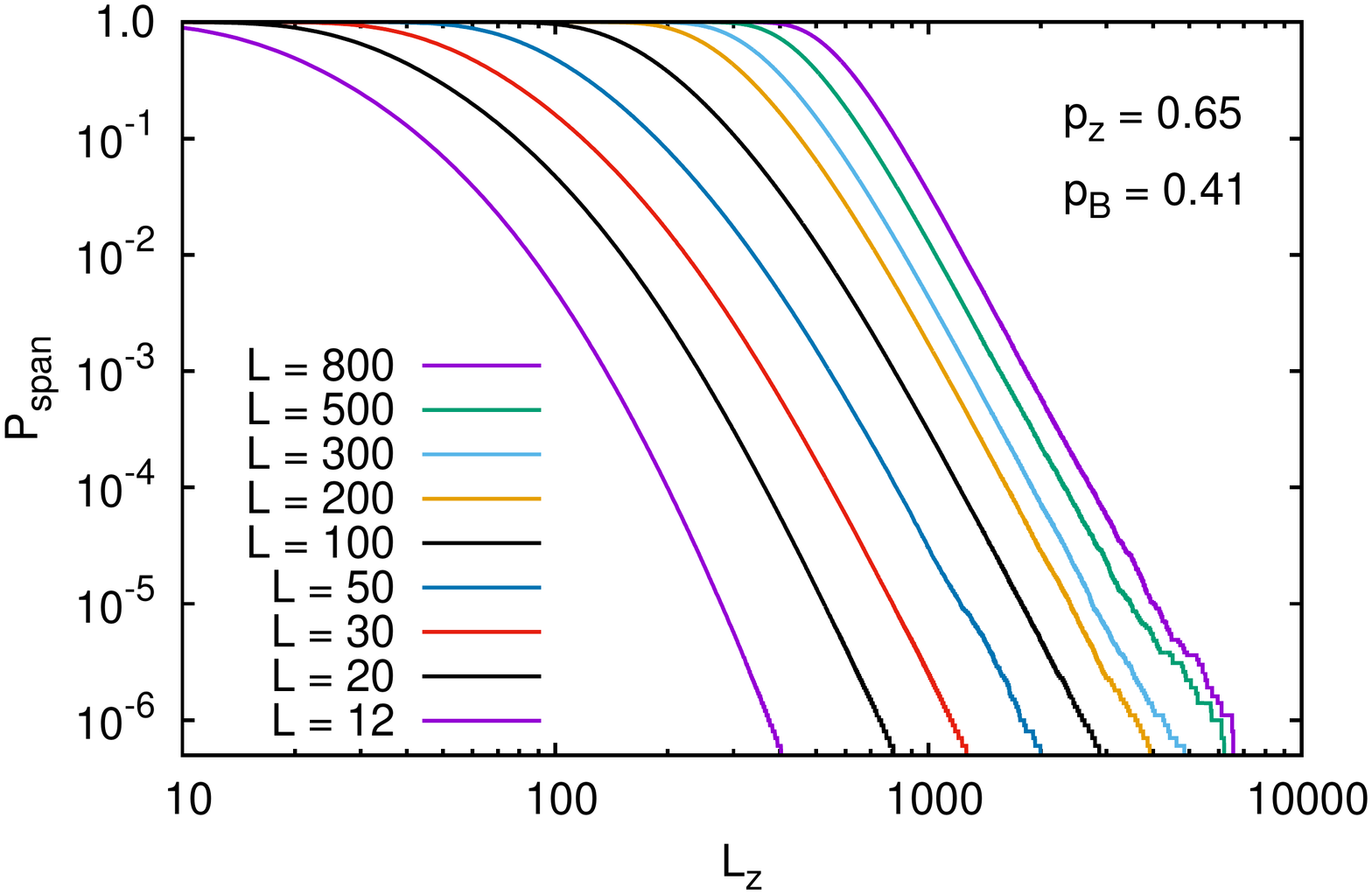}
\vglue -1.cm
\includegraphics[scale=0.30]{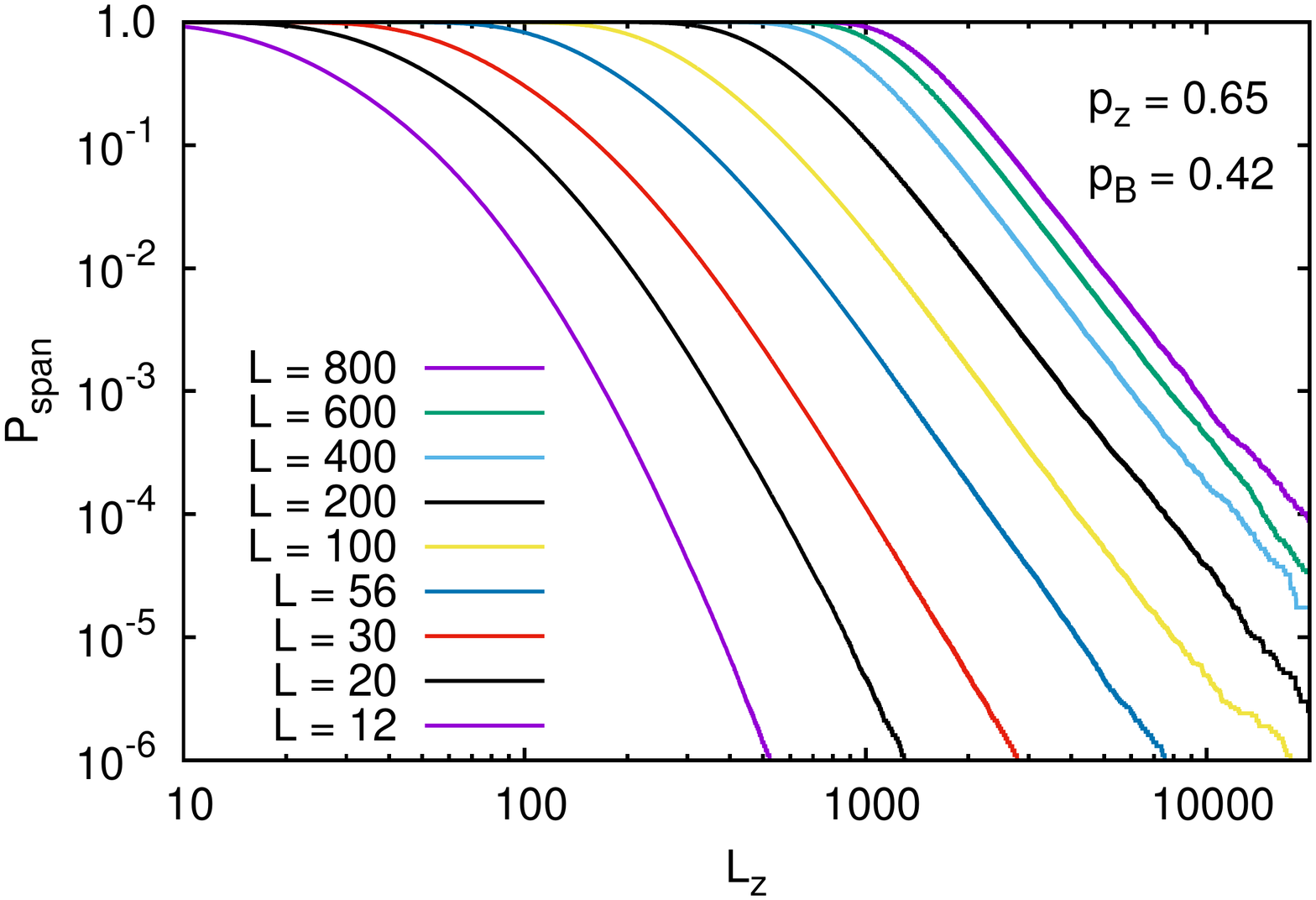}
\par\end{centering}
\caption{\label{fig8} (color online) Log-log plots of spanning probabilities at $p_z = 0.65$,
   plotted against $L_z$. Each curve corresponds to a fixed width $L$. Plot (a) is for $p_B=0.41$, while
   (b) is for $p_B=0.42$.}
\end{figure}

The bound \eqref{Ps} clearly shows that, for the region $p_c^{(3)} < p_B < p_{B,c}(p_z)$, the system is in
a Griffiths phase \cite{Griffiths,Dhar,Moreira,Cafiero}. It mimics very closely the derivation 
for the analogous bound in the case of directed percolation with columnar disorder \cite{Moreira}, 
or equivalently the contact process with frozen disorder. This should not be a surprise. 
Since the columnar disorder and the boundary conditions constrain the main direction of growth 
to be the positive $z$ direction, the main difference between directed and 
undirected percolation is effectively lost. 
Notice that the analogy with directed percolation only holds in the subcritical 
region, but not on the critical line. There, lateral and backward growth is non-negligible, and the 
behavior of critical directed percolation is quite different from the present model.

On the other hand, the bound is also the same as in the drilling percolation problem of \cite{Schrenk}.
The proof uses essentially the arguments, except for the fact that the area $\cal A$ had to be a narrow 
strip along the diagonal in \cite{Schrenk}. This was necessary because only in this way the transversely 
drilled cylinders correspond to short range disorder within $\cal A$. In the present 
case, since the transversely drilled cylinders are replaced by point defects, such a caveat is not needed.
The above theoretical argument leading to the bound \eqref{Ps} is fairly natural, however a completely rigorous proof requires more work.
We present it in the appendix for the interested readers.

In order to test the bound (\ref{Ps}), and to see whether it is saturated already at presently reachable
lattice sizes, we made extensive simulations at $p_z = 0.65$. Results for $p_B = 0.41$ and $0.42$ are shown in 
Fig.~8, where $P_{\rm span}$ is plotted against $L_z$ for various values of $L$. For $L\geq 50$ we see
in both plots very clear power laws $P_{\rm span} \sim L_z^{-\alpha}$ with $\alpha = 5.7(2)$ for $p_B = 0.41$ 
and $\alpha = 3.6(1)$ for $p_B = 0.42$. The power law does not hold for $L= 12$ and $L>100$ (in both panels), 
either because the correlation length in 3-d percolation at $p_B\leq 0.42$ 
is roughly of order 10 to 20, or because Eq.~(\ref{A}) is violated. The fact that the violation of Eq.~(\ref{Ps})
is bigger in panel (b) than in (a), although $\xi$ is smaller for $p_B=0.42$ than for 0.41, indicates that
the latter is the reason. Indeed, Eq.~(\ref{A}) tells us that 
the power law must break down for every fixed $L$, if $L_z$ becomes too large.

\subsection{Subcritical scaling of $P(t)$}

\begin{figure}
\begin{centering}
\includegraphics[scale=0.30]{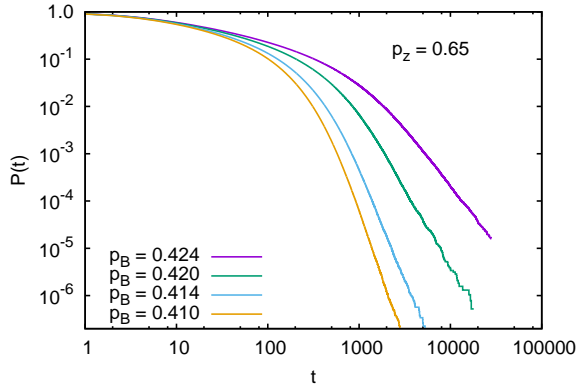}
\par\end{centering}
\caption{\label{fig9} (color online) Log-log plots of $P(t)$ at  $p_z = 0.65$,
   plotted against $t$.}
\end{figure}

Finally, we show in Fig.~9 results for $P(t)$ (which is also equal to the size distribution as measured by
the ``chemical distance" $t$) in the subcritical (Griffiths) phase. We again show data only for $p_z = 0.65$,
but analogous results were seen also at other $p_z$. We clearly see that $P(t)$ decays for large $t$ according
to power laws, where the power depends both on $p_B$ and on $p_z$. The reason for this is of course the same
as in the previous subsection, and the analytic proof should follow along the same lines.

\section{Conclusions}

We showed that replacing some of the point defects in a percolating random material by parallel columnar 
defects changes dramatically the behavior of the percolation transition. Notice that such materials appear 
naturally in many contexts, e.g. by irradiation with energetic radioactive rays or by very controlled 
surface deposition. Clusters become very much 
elongated in the direction of the columns, both at the critical point and below. When treated as an
epidemic growth process, the extension of clusters in the direction parallel to the columns seems to 
grow linearly with time. On the other hand, the extension in the perpendicular direction seems to grow 
only $\sim \sqrt{t}$. Strangely, this dramatic change from OP is not reflected in the growth of the mass 
of critical clusters, which seems to follow exactly the same scaling law as in OP -- except at the 
end point of the transition curve where the columnar defects are strongest and where the scaling changes
abruptly. 

Both heuristically and mathematically this behavior can be understood by viewing the subcritical phase 
as a Griffiths phase, where just the randomness is not ``frozen" in time but is ``frozen" in $z$-direction.
This makes it analogous to the Griffiths phase in directed percolation which can be either understood as 
a purely geometric problem in $d+1$ dimensions of space or as a dynamic problem (the `contact process' or 
SIS epidemics) in $d$ dimensions of 
space. Thus the contact process with frozen disorder can be viewed either as a Griffiths problem in 
the original sense or as a spatial Griffiths problem as in the present paper.

On the other hand we showed that drilling percolation as treated in \cite{Schrenk,Grassberger2016} is 
very similar, and we argued that the power law behaviors in the subcritical phase found mathematically in 
\cite{Schrenk} are also manifestations of a Griffiths phase. If this is true, we might expect that 
the extreme anisotropy of critical clusters found in the present paper should also be seen asymptotically
in drilling percolation. The fact that they are not (yet) seen might then suggest that the true 
asymptotic behavior of drilling percolation has not yet been observed.

\section{Appendix}
This appendix is devoted to provide a completely rigorous proof of bound \eqref{Ps}.
It will roughly follow the same lines as the derivation provided in Section V.
The argument is divided into two main steps: First we choose a `seed' of area $A$ on the plane $z=0$ composed of sites that are not touched by any removed column.
Next we will show that the probability of finding a path above this seed starting from the plane $z=0$ and extending vertically up to height $z=c \exp{(c^{-2}A)}$ is bounded from below by a constant $\delta >0$, uniformly in $A$.
Here, $c>0$ is a positive integer constant whose value is going to be fixed later.
The important fact is to notice that the height $z$ of the spanning path is exponentially larger than the area of the seed.
Therefore, fixing a seed whose area is logarithmically small in comparison with the size of the lattice, allows us to find a path that traverses the lattice with good probability.
Also the probability to find a suitable seed, which is exponentially small in $A$, is then a power of the lattice
size. Together, these two -- the probability to find a seed and the probability to find a path, given a seed -- will
give \eqref{Ps}.

Let us assume from now on that the dimensions of the lattice are $L_x=L_y=L_z=L$ where $L$ is a positive integer that we assume to be large. To start, we choose two integers $c>0$ and $n>0$ that are allowed to grow with $L$
beyond any limit,
as long as $cn \leq L$.
In addition we also fix a seed located in the intersection of the lattice with the plane $z=0$.
We assume that this seed consists of a rectangular strip of thickness $c>0$ and length $c\log(n)$, so that $A = c^2 \log(n)$ (latter we will comment on why we chose this specific restriction for the shape of the seed).
We then ask for the probability that at least one `good' path exists which spans from height 0 to $z=c \exp{(c^{-2}A)} = cn \leq L$.

Now denote by $\mathcal{S}_{B,z}(n)$ the event that there exists such a path $\gamma$ that satifies:
\begin{enumerate}
\item $\gamma$ does not contain any site that has been deleted by the removal of columns;
\item $\gamma$ also does not contain any site that has been deleted by the Bernoulli percolation;
\item $\gamma$ is contained in the portion of the lattice that projects onto the seed in the plane $z=0$;
\item $\gamma$ starts at the seed ($z=0$) and extends up to height $z=cn$.
\end{enumerate}

Similarly, we denote by $\mathcal{S}_B (n)$ the event that there exists a path satisfying conditions 2 to 4, but 
not necessarily 1 (i.e., the path can contain sites in columns that have been drilled).
Finally, we will also need to consider the event $\mathcal{A}_z(n)$ that none of the sites in the seed have been drilled in the $z$-direction.

Note that $\mathcal{S}_B(n)$ only depends on the Bernoulli percolation procedure while $\mathcal{A}_z$ only depends on the columnar mechanism of removal.
Therefore they are independent events.
Furthermore the occurrence $\mathcal{S}_{B,z}(n)$ is assured by the simultaneous occurrence of $\mathcal{S}_B(n)$ and $\mathcal{A}_z(n)$, therefore, for all $n \geq 1$,
\begin{equation}
\label{eq:Prob_S_B,z}
\begin{split}
\mathbb{P}_{p_B,p_z} (\mathcal{S}_{B,z}(n)) & \geq \mathbb{P}_{p_z} (\mathcal{A}_z(n))\mathbb{P}_{p_B} (\mathcal{S}_B(n)) \\ & = p_z^{c^2 \log(n)} \mathbb{P}_{p_B}(\mathcal{S}_B(n)).
\end{split}
\end{equation}

Assume now that there exist $c>0$ and $\delta>0$ such that
\begin{equation}
\label{eq:Prob_S_B}
\mathbb{P}_{p_B} (\mathcal{S}_B(n)) \geq \delta,
\end{equation}
for all $n \geq 1$.

Plugging into \eqref{eq:Prob_S_B,z} we get:
\begin{equation}
\label{eq:Prob_Span}
\mathbb{P}_{p_B, p_z} (\mathcal{S}_{B,z}(n)) \geq p_z^{c^2 \log(n)} \delta = \delta n^{-\alpha}.
\end{equation}
for all $n \geq 1$.
For the special case $n = L/c$ we obtain:
\begin{equation*}
\label{eq:Prob_Span_L}
\mathbb{P}_{p_b,p_z} 
\left(
\begin{gathered}
\text{there is an open path $\gamma$}\\
\text{spanning the lattice vertically}
\end{gathered}
\right) \geq ({\delta}{c^{\alpha}}) L^{-\alpha}
\end{equation*}
which is exactly Eq.\ \eqref{Ps}.

From the discussion above, in order to conclude the proof, it is sufficient to find $c>0$ and $\delta>0$ for which \eqref{eq:Prob_S_B} holds for all $n \geq 1$.
Before we tackle this problem, let us mention that the exponent $\alpha$ above depends on both  $c$ and $p_z$.
Since the value of $c$ to be fixed latter will depend on $p_B$, we conclude that $\alpha$ actually depends on both $p_z$ and $p_B$.

Let us now move to the proof of \eqref{eq:Prob_S_B}.
The main technique we use is the so-called \textit{one-step renormalization} or \textit{block argument}.
Roughly speaking it consists of tiling the lattice with cubes of side length $c$ (called blocks) and then working on a new \textit{renormalized lattice} where the role of the sites are played by the blocks and where two blocks are considered adjacent (neighbors) whenever they share a face.

Let us denote by $S(c, n)$ the slab shaped region consisting of all the sites in the lattice located above the fixed seed and whose height range from $z=0$ to $z=cn$.
Notice that the corresponding region in the renormalized lattice is just an $n \times \log(n)$ rectangle, thus it is strictly two-dimensional. (That's the reason why we have picked the strip-shaped seed. Other choices would have given a more complicated region.) 

For a particular block $B$ in $S(c,n)$, typically, there are $8$ other blocks in $S(c,n)$ sharing a face or a line segment with $B$.
Define $\tilde{B}$ as being the union of $B$ and these $8$ blocks:
\[
\tilde{B} = \bigcup_{j,k \in \{0,\pm 1\}} B+jc e_2 + kc e_3,
\]
where $e_2$ and $e_3$ stand for the unit vectores in the $z$ and $y$ direction. (In the case that $B$ does not lie in the bulk of $S(c,n)$ there will be less neighbors, however, the arguments we present go along the same lines.)

For a fixed block $B$ we say that the event $\mathcal{U}(B)$ occurs if the Bernoulli site percolation process restricted to $\tilde{B}$ satisfies:
\begin{enumerate}
\item There exists a unique cluster $C(B) \subset \tilde{B}$ with (maximum norm) diameter greater or equal to $3c$.
\item The cluster $C(B)$ intersects every cube of side length $c$ contained in $\tilde{B}$.
\item The cluster $C(B)$ touches all the faces of $\tilde{B}$.
\end{enumerate}
\begin{definition}
When the event $\mathcal{U}(B)$ occurs we say that $B$ is a well-connected block, and $C(B)$ is ``spanning" $\tilde{B}$.
\end{definition}

Notice that the occurrence of event $\mathcal{U}(B)$ requires the cluster $C(B)$ to be unique. Although $B$ might seem even 
better connected if more than one spanning cluster occurs in $\tilde{B}$, this would not be sufficient for the following 
arguments. 
The occurrence of event $\mathcal{U}(B)$ implies that a good portion of $\tilde{B}$ is occupied by $C(B)$.
Also, as we will show below, it follows from its definition that the spanning clusters of two adjacent well-connected blocks will be connected. 
In the following, this heuristic argument will be made precise.

Indeed, the assumption $p_B > p_c(\mathbb{Z}^3)$ guarantees that in the limit that $c \to \infty$ this will be true with 
overwhelming probability. This is a straightforward fact in supercritical Bernoulli percolation that we summarise
as follows:
\begin{proposition}
\label{prop:block_well_conn}
For any fixed $p_B > p_c^{(3)}$,
\begin{equation}
\lim_{c \to \infty} \mathbb{P}_{p_B}
\left(
\begin{gathered}
\text{a block $B$ of side-length $c$}\\
\text{is well-connected}
\end{gathered}
\right) =1 .
\label{eq:prob_well_conn}
\end{equation}
\end{proposition}
We refer the reader to either \cite[Theorem 3.1]{Pisztora} or \cite[Theorem 5]{Pisztora_Penrose} for a rigorous proof of this proposition.
There the authors provide quantitive lower and upper bound estimates for the rate of convergence in \eqref{eq:prob_well_conn}.
\medskip

Our next goal is to show how to use well-connected blocks in order to create long open paths inside $S(c,n)$.
Since $S(c,n)$ corresponds in the renormalized lattice to a $\log(n) \times n$ rectangle,  one can think of the configuration of well-connected blocks as the realization of a $2$-d percolation model in this renormalized lattice.
This is not an independent percolation as the state of each block depends on the state of its imediate neighboring blocks.
However the dependencies are only finite range.
Indeed the events $\mathcal{U}(B_1)$ and $\mathcal{U}(B_2)$ are independent as soon as $\text{dist}(B_1,B_2) >2c$.

We now claim that
\begin{equation}
\label{eq:glue}
\begin{gathered}
\text{If $B_1$ and $B_2$ are two neighboring well-} \\ \text{connected blocks, then $\mathcal{C}(B_1)\cap \mathcal{C}(B_2) \neq \emptyset$.}
\end{gathered}
\end{equation}
Notice that the above claim is purely deterministic.
In fact, it is a direct consequence of the geometry involved in the definition of the events $\mathcal{U}(B_1)$ and $\mathcal{U}(B_2)$ as we show next:
Assume that $B_1$ and $B_2$ are neighbours and that $\mathcal{U}(B_1)$ and $\mathcal{U}(B_2)$ occurs.
For simplicity let us also assume that $B_1$ and $B_2$ are in the bulk of the slab shaped region $S(c,n)$ so that the region $\tilde{B}_1 \cap \tilde{B}_2$ comprises $6$ cubes of side-length $c$ from used in the paving of $\mathbb{Z}^3$.
The occurrence of $\mathcal{U}(B_1)$ guarantees that $C(B_1)$ has to intersect all of these $6$ cubes.
From this we conclude that $C(B_1) \cap \tilde{B}_2$ contains at least one cluster of (maximum norm) diameter greater than $c$.
Now the uniqueness of $C(B_2)$ required in the definition of the event $\mathcal{U}(B_2)$ assures that any cluster of diameter greater than $c$ in $\tilde{B}_2$ must be contained in $C(B_2)$.
Therefore, $C(B_1) \cap C(B_2) \neq \emptyset$ which proves \eqref{eq:glue}.

In fact, since ${C}(B_1)$ has to touch every cube of side length $c$ contained in $\tilde{B}_1$, ${C}(B_1) \cap \tilde{B}_2$ has diameter greater than $c$.
By the uniqueness of ${C}(B_2)$ inside $\tilde{B}_2$ it has to contain ${C}(B_1) \cap \tilde{B}_2$.

The above claim provides a handy way of gluing the clusters of neighboring well-connected blocks.
Indeed, if we find a sequence of neighboring well-connected blocks, \eqref{eq:glue} then the clusters inside each one of them are part of a larger cluster that extends inside the union of all of them.
Since each of the clusters touch the faces of the corresponding block, we can navigate inside this sequence of blocks passing through a path of open sites.
We state this as a proposition:
\begin{proposition}
\label{prop:path}
If there exists a path $B_1, B_2, \ldots, B_k$ of successive neighboring well-connected blocks spanning the renormalized region $S(c,n)$ from top to bottom, then there existis a path $\gamma$ of open sites such that
\[
\gamma \subset [\mathcal{C}(B_1) \cup \mathcal{C}(B_2) \cdots \cup \mathcal{C}(B_k)] \cap S(c,n)
\]
and such that $\gamma$ spans $S(c,n)$ from top to bottom.
In particular, the event $\mathcal{S}_B(n)$ occurs.
\end{proposition}

The idea now is to tune $c$ in order to show that a sequence of well-connect blocks can be found with good probability inside the slab $S(c,n)$.
For that we first recall that $S(c,n)$ corresponds in the renormalized lattice to a $n\times \log(n)$ rectangle  where the process of well-connected blocks presents finite range dependencies.
Second, we state the following straightforward fact: For $2$-d Bernoulli percolation, the probability of spanning the rectangle $[0, \log{n}] \times [0,n]$ converges to $1$ as $n \to \infty$ provided that the retention parameter $p$ is large enough, say $p \geq p_o$ for some $p_o > p_c^{(2)}$ large enough \cite[Theorem 11.55]{Grimmett}.
The same remains true when Bernoulli percolation is replaced by a given finite-range dependent percolation  \cite[Theorem 0.0]{Liggett} (maybe increasing the value of $p_o$ accordingly).

\begin{proposition}
\label{prop:block} There exists a $\delta>0$ and $p_\circ \in ]p_c^{(2)}, 1[$ such that, if
\[ 
\mathbb{P}_{p_B}
\left(
\begin{gathered}
\text{a block $B$ is well-connected }
\end{gathered}
\right) \geq p_\circ 
\]
then
\begin{equation}
\mathbb{P}_{p_B}
\left(
\begin{gathered}
\text{there exists a path of well-connected}\\
\text{blocks spanning $S(c,n)$ vertically}
\end{gathered}
\right) > \delta.
\end{equation}
\end{proposition}

It is now clear, from the Propositions \ref{prop:path} and \ref{prop:block} above, that all we need to do in order to obtain Eq.\ $\eqref{eq:Prob_S_B}$ is to prove that if we chose $c$ very large, than a supercritical percolation process restricted to a cube $B$ of side-length $c$ will fulfil the three items in the definition of the event $\mathcal{U}(B)$ with very high probability (eventually larger than $p_\circ$).
In view of Proposition \ref{prop:block_well_conn}, there exists $c$ depending on $p_\circ$ that fulfils this condition.
Therefor, putting together the three previous propositions readily implies that Eq.\ \eqref{eq:Prob_S_B} holds for all $n$ provided that the value of $c$ is chosen sufficiently large.

\noindent \textbf{Acknowledgements.}
We thank Julian Schrenk, Nuno Ara\'ujo, and Hans Herrmann for stimulating correspondence. Marcelo Hil\'ario thanks NYU Shanghai for its hospitality.

\end{document}